\documentclass[aps,prl,reprint,amsmath,amssymb,superscriptaddress,nofootinbib,preprintnumbers]{revtex4-2}

\usepackage{graphicx}
\usepackage{mathtools}
\usepackage{microtype}
\usepackage{placeins}
\usepackage{capt-of}
\usepackage[hidelinks]{hyperref}
\usepackage{xcolor}

\hypersetup{
  pdftitle={Stokes Phenomena between AdS/CFT and dS/CFT},
  pdfauthor={Masazumi Honda and Kotaro Shinmyo}
}

\newcommand{\order}{\mathcal O}
\newcommand{\Log}{\operatorname{Log}}

\begin{document}

\preprint{RIKEN-iTHEMS-Report-26, STUPP-26-301, YITP-26-123}

\title{Stokes Phenomena between AdS/CFT and dS/CFT}

\author{Masazumi Honda}
\email{masazumi.honda@riken.jp}
\affiliation{RIKEN Center for Interdisciplinary Theoretical and Mathematical Sciences (iTHEMS), Wako, Saitama 351-0198, Japan}
\affiliation{Graduate School of Science and Engineering, Saitama University, Saitama 338-8570, Japan}

\author{Kotaro Shinmyo}
\email{kotaro.shinmyo@yukawa.kyoto-u.ac.jp}
\affiliation{Yukawa Institute for Theoretical Physics, Kyoto University, Kyoto 606-8502, Japan}

\date{\today}

\begin{abstract}
As parameters are varied, the set of saddle points contributing to a (path) integral may change discontinuously, leading to a corresponding change in the asymptotic expansion of the integral.
This behavior is known as the Stokes phenomenon.
We explore this phenomenon in the context of the analytic continuation problem relating the AdS/CFT and dS/CFT correspondences.
In this paper, we study these correspondences for three-dimensional pure gravity and two-dimensional Liouville theory, using independent calculations in bulk minisuperspace and in the boundary Liouville zero-mode.
In the bulk, the dS contour selects a single saddle and yields the tunneling wave function.
Upon continuation to AdS, the contour instead selects an infinite family of saddles.
The boundary calculation independently reproduces the same Stokes structure at leading semiclassical order, providing a nontrivial holographic consistency check.
Our construction also offers a contour prescription for the conformal factor problem in Euclidean AdS$_3$ quantum gravity within minisuperspace.

\end{abstract}

\maketitle

%%%%%%%%%%%%%%%%%%%%%%%%%%%%%
%%%%%%%%%%%%%%%%%%%%%%%%%%%%%
%%%%%%%%%%%%%%%%%%%%%%%%%%%%%
\section{Introduction}
%%%%%%%%%%%%%%%%%%%%%%%%%%%%%
%%%%%%%%%%%%%%%%%%%%%%%%%%%%%
%%%%%%%%%%%%%%%%%%%%%%%%%%%%%

A nonperturbative definition of quantum gravity remains a central problem in theoretical physics.
A natural starting point is the gravitational path integral.
Even if we set aside UV divergences and the problem of defining the path-integral measure, we must still specify an integration contour in the space of fields, including the metric.
This issue is especially acute in Euclidean gravity because the Euclidean Einstein--Hilbert action is unbounded below along the conformal mode.
As a result, the integral over real Euclidean metrics does not converge.
This differs from ordinary quantum field theory, where Wick rotation usually improves convergence.
The resulting contour-selection issue arises both in gravitational partition functions \cite{Gibbons:1976ue,Hawking:1978jz,Gibbons:1978ac} and in proposals for the wave function of the universe \cite{Hartle:1983ai,Vilenkin:1982de,Vilenkin:1986cy,Linde:1983mx}.

It is generally difficult to determine an appropriate integration contour for the gravitational path integral without knowing what quantum gravity actually is.
However, it may still be possible to draw some lessons from analyzing naive path integral quantization of general relativity. 
In this paper, we use analytic continuation and holographic consistency to constrain the integration contour, and Picard--Lefschetz (PL) theory to determine the contributing saddles.

Picard--Lefschetz theory provides a systematic framework for analyzing such contour choices and determining which saddle points contribute to an integral \cite{Witten:2010cx}.
Its application to cosmological path integrals has been developed in \cite{Halliwell:1988ik,Feldbrugge:2017kzv,DiazDorronsoro:2017hti,Honda:2024aro,Ailiga:2025fny}.
For an integral
\begin{align}
 Z(\lambda)
 &=\int_{\mathcal C} d^n z\, e^{F(z;\lambda)},
 \notag\\[-1mm]
 \mathcal C
 &\sim \sum_\sigma n_\sigma \mathcal J_\sigma,
 \qquad n_\sigma\in\mathbb Z,
 \label{eq:PL-intro}
\end{align}
the Lefschetz thimble $\mathcal J_\sigma$ is a steepest-descent cycle passing through a saddle point $z_\sigma$ satisfying $\partial_iF(z_\sigma;\lambda)=0$.
The coefficient $n_\sigma$ determines whether that saddle contributes.
As $\lambda$ is varied, the coefficients $n_\sigma$ can jump discontinuously.
The integral itself does not jump, but its decomposition into saddle-point contributions changes.
This is the Stokes phenomenon \cite{Pham:1983,Berry:1991but,Howls:1997}.

A complementary approach to quantum gravity is provided by holography \cite{tHooft:1993dmi,Susskind:1994vu}. 
Its best-established realization is the AdS/CFT correspondence, which relates gravitational theories in asymptotically AdS spacetime to conformal field theories \cite{Maldacena:1997re,Gubser:1998bc,Witten:1998qj}.
Motivated by this success, holographic descriptions of de Sitter space have been proposed in \cite{Strominger:2001pn,Witten:2001kn}.
Analytic continuation from Euclidean AdS provides a related way to define dS wave functions and correlators \cite{Maldacena:2002vr,Castro:2012gc,Cotler:2019nbi}.
%Three-dimensional gravity 
Three-dimensional pure gravity is a particularly tractable setting for these ideas as there are no local propagating degrees of freedom.
For negative cosmological constant, Einstein gravity admits a non-chiral Chern--Simons formulation \cite{Achucarro:1986uwr,Witten:1988hc,Cacciatori:2001un}.\footnote{
Virasoro TQFT quantizes Teichm\"uller space and has been proposed as a framework for pure (A)dS$_3$ quantum gravity \cite{Collier:2023fwi,Collier:2024mgv,Collier:2025lux}.
Its precise relation to the Liouville holographic setup and integration contours considered here remains to be clarified.
}
This formulation induces a boundary non-chiral WZW model \cite{Witten:1988hf,Elitzur:1989nr,Coussaert:1995zp,
Cacciatori:2001un}, and Brown--Henneaux boundary conditions further lead to Liouville theory \cite{Brown:1986nw,Coussaert:1995zp,Drinfeld:1984qv,Forgacs:1989ac,Balog:1990mu,Cacciatori:2001un,Takhtajan:2002cc}:
\begin{equation}
 3d\;SL(2,\mathbb R)\;\mathrm{CS}
 \longleftrightarrow 2d\;SL(2,\mathbb R)\;\mathrm{WZW}
 \longrightarrow 2d\;\mathrm{Liouville}.
 \label{eq:CS-WZW-L}
\end{equation}
Complex Liouville theory has been proposed as a holographic description of $\mathrm{dS}_3$ gravity \cite{Hikida:2021ese,Hikida:2022ltr}.
The proposal has been developed through bulk correlators, late-time CFT correlators, and analyses of complex Chern--Simons and metric saddles \cite{Chen:2022ozy,Chen:2022xse,Chen:2023prz,Chen:2023sry,Chen:2024qmn,Chen:2024vpa}.
At leading semiclassical order, the central charge $c$ of the Liouville theory is related to the (A)dS radius $\ell_{\rm (A)dS}$ by $c=\frac{3\ell_{\mathrm{AdS}}}{2G_N}$ for AdS$_3$ and $c=i\frac{3\ell_{\mathrm{dS}}}{2G_N}$ for dS$_3$.

This correspondence leads to a natural consistency condition.
If holography is to hold, the bulk and boundary path integrals should select compatible integration cycles.
Furthermore, the set of contributing saddles should exhibit compatible changes under analytic continuation of parameters, such as $\frac{\ell_\mathrm{(A)dS}}{G_N}$ or $c$.

In this paper, we test this condition by comparing the bulk and boundary theories in their spherically symmetric sectors.
In the bulk, we define the dS$_3$ minisuperspace path integral from the positive Lorentzian lapse, determine its Lefschetz thimble decomposition, and then define the AdS$_3$ contour by analytically continuing the radius $\ell_\mathrm{(A)dS}$ and the bulk cutoff parameter $a_1$.
On the boundary, we reduce Liouville theory to its constant mode.
Starting from a convergent Lefschetz thimble, we define the AdS and dS integrals by analytically continuing the central charge $c$ and the boundary cutoff parameter $\epsilon$.

Both calculations independently reduce to Gamma functions and exhibit the same Stokes phenomenon in their saddle-point decompositions under analytic continuation between dS and AdS.
This agreement provides a nontrivial consistency check of holography.
The dS contour is homologous to a single thimble passing through the tunneling saddle, with
$|\Psi_{\mathrm{dS}}|\sim\exp(-\frac{\pi\ell_{\mathrm{dS}}}{4G_N})$,
whereas the AdS contour is homologous to an infinite sum of thimbles.

%%%%%%%%%%%%%%%%%%%%%%%%%%%%%
%%%%%%%%%%%%%%%%%%%%%%%%%%%%%
%%%%%%%%%%%%%%%%%%%%%%%%%%%%%
\section{Three-dimensional minisuperspace gravity}
%%%%%%%%%%%%%%%%%%%%%%%%%%%%%
%%%%%%%%%%%%%%%%%%%%%%%%%%%%%
%%%%%%%%%%%%%%%%%%%%%%%%%%%%%

Our bulk analysis starts from the dS integration cycle defined by the positive Lorentzian lapse $N\in i\mathbb R_+$.
The dimensionless bulk cutoff parameter $a_1$ determines the holographic UV cutoff length $\frac{\ell}{a_1}$ \cite{Susskind:1998dq,Ryu:2006ef,Heemskerk:2010hk,Casini:2011kv}.
We keep this length fixed and positive while analytically continuing $\ell$ and $a_1$ to the AdS parameters.
We then determine an AdS lapse contour that reproduces this continuation.

To implement this prescription, consider the Einstein--Hilbert action on a three-ball,
\begin{equation}
 I_{\rm grav}=-\frac1{16\pi G_N}\int\!d^3x\sqrt g\,(R-2\Lambda)
 +I_{\rm GH}+I_{\rm ct},
 \label{eq:EH}
\end{equation}
and the spherically symmetric metric
\begin{equation}
 \begin{aligned}
 ds^2&= \ell^2\left(N(\tau)^2d\tau^2+a(\tau)^2d\Omega_2^2\right),\\[-1mm]
 a(0)&=0,
 \qquad a(1)=a_1.
 \end{aligned}
 \label{eq:metric}
\end{equation}
After fixing $N'=0$, the reduced path integral is
\begin{align}
 Z_{\mathrm{AdS}} \ \text{or}\ \Psi_{\mathrm{dS}}
 &=\int_{\mathcal C}\!dN\int\!\mathcal Da\,
  e^{-I[a;N]-I_{\rm ct}},
 \notag\\[-1mm]
 I[a;N]
 &=-\frac{\ell_{(A)\mathrm{dS}}}{2G_N}\int_0^1\!d\tau\,N
 \left[N^{-2}(a')^2\pm a^2+1\right],
 \label{eq:mini-action}
\end{align}
where the upper and lower signs refer to AdS and dS, respectively.
At finite $a_1$, evaluating the Gaussian path integral over $a(\tau)$ at fixed $N$ and fixing its normalization by the short-time limit gives, up to an $N$-independent constant,\footnote{
The prefactors in \eqref{eq:ZA-lapse} and \eqref{eq:Psi-lapse} differ slightly from those in \cite{Chen:2024qmn,Chen:2024vpa}.
We obtain this prefactor from the standard textbook result for the harmonic-oscillator propagator \cite{Feynman:100771,Mukhanov:2007zz}.
Its application in the present quantum-cosmology setting is discussed in \cite{Honda:2026wsl}.
}
\begin{align}
 Z_{\mathrm{AdS}}
 &=\int_{\mathcal C}\frac{dN}{\sqrt{\sinh N}}
 e^{\frac{\ell_{\mathrm{AdS}}}{2G_N}
 \left(N+a_1^2\coth N\right)-I_{\rm ct}^{\mathrm{AdS}}},
 \label{eq:ZA-lapse}\\
 \Psi_{\mathrm{dS}}
 &=\int_{\mathcal C}\frac{dN}{\sqrt{\sin N}}
 e^{\frac{\ell_{\mathrm{dS}}}{2G_N}
 \left(N+a_1^2\cot N\right)-I_{\rm ct}^{\mathrm{dS}}}.
 \label{eq:Psi-lapse}
\end{align}
The same minisuperspace action and its semiclassical saddle geometries were studied in \cite{Chen:2024qmn,Chen:2024vpa}.

%%%
\begin{figure*}[t]
 \centering
 \makebox[\textwidth][c]{%
  \includegraphics[width=0.3\textwidth]{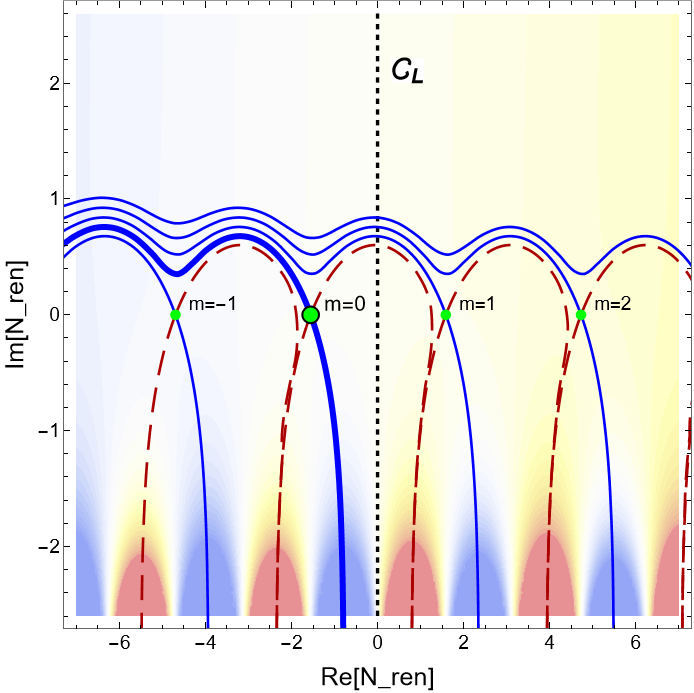}%
  \hspace{0.1\textwidth}%
  \includegraphics[width=0.3\textwidth]{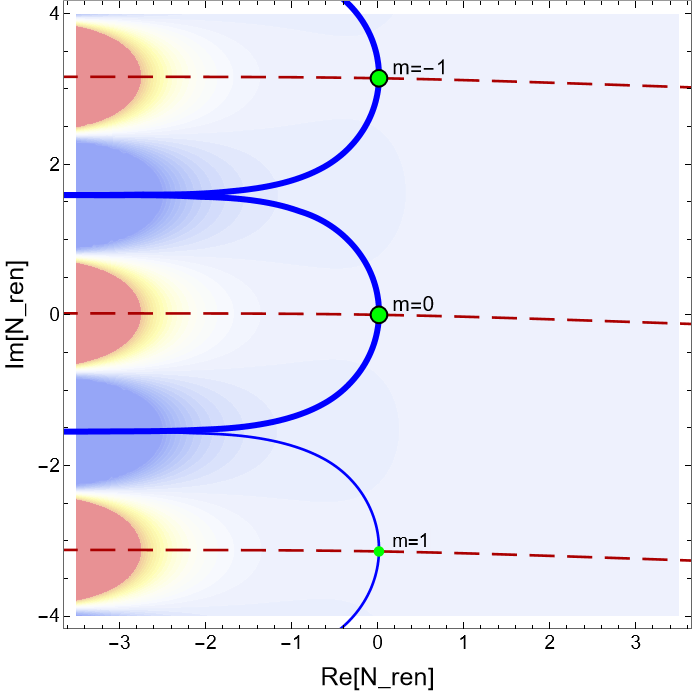}%
 }
\caption{
Lefschetz-thimble decompositions in the complex $N_{\rm ren}$ plane for dS (left) and AdS (right) minisuperspace gravity.
The blue curves and red dashed curves denote the downward thimbles $\mathcal J_m$ and upward cycles $\mathcal K_m$, respectively, while the green points denote the saddles $N_m$.
Contributing thimbles and saddles are emphasized by thicker blue curves and green circles.
Left: the Lorentzian dS integral with $\ell_{\mathrm{dS}}/(4G_N)=8$.
The original contour $\mathcal C_{\rm L}$ is homologous to $\mathcal J_0$.
Right: the AdS integral with $\frac{\ell_{\mathrm{AdS}}}{4G_N}=\sqrt{8^2-0.3^2}+0.3i\simeq7.99437+0.3i$,
so that its absolute value is also $8$ and the Gamma-function argument
$1/4-\ell_{\mathrm{AdS}}/(4G_N)$ lies below the negative real axis.
The continued contour is homologous to
$\mathcal J_0+\mathcal J_{-1}+\mathcal J_{-2}+\cdots$.
}
 \label{fig:bulk-thimbles}
\end{figure*}
%%%

To keep track of the cutoff dependence, we choose the dS counterterm branch $I_{\rm ct}^{\mathrm{dS}}=-2i\frac{\ell_{\mathrm{dS}}}{4G_N}a_1^2$ and define $N_{\rm ren}=N-i\log(2a_1)$, where the branch of $\Log$ is continued from $a_1>0$.
In the large-$a_1$ limit, \eqref{eq:Psi-lapse} becomes, up to an $a_1$-independent normalization,
\begin{align}
 \Psi_{\mathrm{dS}}
 &\propto(2a_1)^{\frac{i\ell_{\mathrm{dS}}}{2G_N}-1/2}
 \int_{\mathcal C_{\rm L}}\!dN_{\rm ren}
 \notag\\[-1mm]
 &\quad\times
 e^{
 2\left(\frac{\ell_{\mathrm{dS}}}{4G_N}+\frac{i}{4}\right)N_{\rm ren}
 -i\frac{\ell_{\mathrm{dS}}}{4G_N}e^{2iN_{\rm ren}}}.
 \label{eq:PsidSren}
\end{align}
The contour in \eqref{eq:PsidSren} is inherited directly from the positive Lorentzian lapse, regulated as $N\in-\varepsilon+i\mathbb R_+$.
At finite $a_1$, the shift $N_{\rm ren}=N-i\log(2a_1)$ maps this contour to
\begin{align}
 \mathcal C_{\rm L}(a_1;\varepsilon)
 &=
 -\varepsilon
 +i\bigl(-\log(2a_1),\infty\bigr).
 \notag
\end{align}
As $a_1\to+\infty$ at fixed $\varepsilon>0$, the lower endpoint moves to $-i\infty$.
Taking $\varepsilon\to0^+$ then gives
\begin{align}
 \mathcal C_{\rm L}
 &=
 \lim_{\varepsilon\to0^+}
 \left(
 -\varepsilon+i\mathbb R
 \right).
 \label{eq:CL-prescription}
\end{align}
Thus, after the cutoff-dependent shift, the positive Lorentzian half-line becomes a full vertical contour.
The small negative real part ensures convergence at the lower end and specifies how the limiting contour is approached.

Having fixed the contour, we now apply the Lefschetz thimble decomposition \eqref{eq:PL-intro} to the dS lapse integral.
Let $F_{\mathrm{dS}}(N_{\rm ren})$ denote the exponent of the integrand in \eqref{eq:PsidSren},
\begin{align}
 F_{\mathrm{dS}}(N_{\rm ren})
 &=
 2\left(
 \frac{\ell_{\mathrm{dS}}}{4G_N}+\frac{i}{4}
 \right)N_{\rm ren}
 -\frac{i\ell_{\mathrm{dS}}}{4G_N}e^{2iN_{\rm ren}}.
 \label{eq:FdS}
\end{align}
Its saddle points satisfy $\partial_{N_{\rm ren}}F_{\mathrm{dS}}(N_m)=0$.
Since they are separated by $\pi$ in the real direction, we label them as
\begin{align}
 N_m
 &=N_0+\pi m,
 \qquad m\in\mathbb Z,
 \label{eq:dS-saddles}
\end{align}
where $N_0$ is the saddle with $-\pi<\operatorname{Re}N_0<0$.

For the integrand $e^{F_{\mathrm{dS}}}$, the downward flow equation is
\begin{align}
 \frac{dN_{\rm ren}}{dt}
 &=
 -\overline{\frac{\partial F_{\mathrm{dS}}}{\partial N_{\rm ren}}},
 \notag\\[-1mm]
 \frac{d}{dt}\operatorname{Re}F_{\mathrm{dS}}
 &=
 -\left|
 \frac{\partial F_{\mathrm{dS}}}{\partial N_{\rm ren}}
 \right|^2,
 \qquad
 \frac{d}{dt}\operatorname{Im}F_{\mathrm{dS}}=0.
 \label{eq:dS-flow}
\end{align}
The Lefschetz thimble $\mathcal J_m$ consists of all downward-flow trajectories that approach the saddle $N_m$ as $t\to-\infty$.
Reversing the sign in the first equation defines the upward cycle $\mathcal K_m$.
Because $\operatorname{Im}F_{\mathrm{dS}}$ is constant along each flow, a flow can connect two saddle points only when the imaginary parts of their saddle-point values agree.

The coefficient $n_m$ of $\mathcal J_m$ in the thimble decomposition is determined by the intersection number of the original integration contour $\mathcal C_{\rm L}$ with $\mathcal K_m$.
For the Lorentzian contour, we obtain
\begin{equation}
 n_m
 =
 \langle\mathcal C_{\rm L},\mathcal K_m\rangle
 =
 \delta_{m0},
 \qquad
 \mathcal C_{\rm L}\sim\mathcal J_0,
 \label{eq:dS-PL}
\end{equation}
as also shown in the left panel of Fig.~\ref{fig:bulk-thimbles}.
Thus only the saddle $N_0$ contributes to the Lorentzian contour, and
\begin{align}
 \Psi_{\mathrm{dS}}
 &\propto
 \left(
 4i\frac{\ell_{\mathrm{dS}}}{4G_N}a_1^2
 \right)^{i\frac{\ell_{\mathrm{dS}}}{4G_N}-\frac14}
 \notag\\[-1mm]
 &\quad\times
 \Gamma\!\left(\frac14-i\frac{\ell_{\mathrm{dS}}}{4G_N}\right)
 \left[1+\order(a_1^{-2})\right],
 \label{eq:Psi-Gamma-cutoff}
\end{align}
where the large-$a_1$ expansion is taken at fixed $\ell_{\mathrm{dS}}/G_N$.\footnote{
The integral \eqref{eq:Psi-lapse} admits a closed-form expression at finite cutoff.
Expanding this exact result for large $a_1$ reproduces \eqref{eq:Psi-Gamma-cutoff};
see the Supplemental Material.
}
Stirling's formula gives\footnote{In the semiclassical formulas below, we suppress cutoff-dependent factors and retain only the leading classical exponential weight of each saddle.}
\begin{align}
 \Psi_{\mathrm{dS}}
 \sim
 e^{
 -\frac{\pi\ell_{\mathrm{dS}}}{4G_N}
 +i\frac{\ell_{\mathrm{dS}}}{4G_N}}.
 \label{eq:bulk-dS}
\end{align}
This is the tunneling wave function, characterized semiclassically by Vilenkin's tunneling boundary condition \cite{Vilenkin:1986cy}.

We obtain the AdS partition function by analytically continuing \eqref{eq:Psi-Gamma-cutoff}.
We rotate $\ell$ and $a_1$ clockwise by $\frac{\pi}{2}$, keeping the cutoff length $\frac{\ell}{a_1}$ fixed and positive and $G_N$ unchanged:
\begin{align}
 \frac{\ell_{\mathrm{dS}}}{4G_N}
 &\longrightarrow
 -i\frac{\ell_{\mathrm{AdS}}}{4G_N},
 \qquad
 a_{1,\mathrm{dS}}
 \longrightarrow
 -ia_{1,\mathrm{AdS}},
 \label{eq:analyticcont}
\end{align}
which gives
\begin{align}
 Z_{\mathrm{AdS}}
 &\propto
 \left(
 -4\frac{\ell_{\mathrm{AdS}}}{4G_N}a_1^2
 \right)^{\frac{\ell_{\mathrm{AdS}}}{4G_N}-\frac{1}{4}}
 \notag\\[-1mm]
 &\quad\times
 \Gamma\!\left(\frac14-\frac{\ell_{\mathrm{AdS}}}{4G_N}\right)
 \left[1+\order(a_1^{-2})\right],
 \label{eq:ZA-Gamma-cutoff}
\end{align}
up to a normalization independent of $a_1$ in the large-$a_1$ limit.

To represent this result as an AdS lapse integral, we choose $I_{\rm ct}^{\mathrm{AdS}}=2\frac{\ell_{\mathrm{AdS}}}{4G_N}a_1^2$, the standard local AdS counterterm \cite{Balasubramanian:1999re}, and introduce $N_{\rm ren}=N-\log(2a_1)$.
We determine $\mathcal C_{\mathrm{AdS}}$ by matching the resulting large-cutoff integral to \eqref{eq:ZA-Gamma-cutoff}, including the prefactor and its continued logarithmic branch.
The exponent of this integral is
\begin{align}
 F_{\mathrm{AdS}}(N_{\rm ren})
 &=
 2\left(
 \frac{\ell_{\mathrm{AdS}}}{4G_N}-\frac14
 \right)N_{\rm ren}
 +\frac{\ell_{\mathrm{AdS}}}{4G_N}e^{-2N_{\rm ren}}.
 \label{eq:FAdS}
\end{align}
In the semiclassical regime, its saddle points are
\begin{align}
 N_m
 &=
 -\frac12\log\left(
 1-\frac{G_N}{\ell_{\mathrm{AdS}}}
 \right)
 -\pi i m,
 \qquad m\in\mathbb Z.
 \label{eq:AdS-saddles}
\end{align}

Applying the flow equation \eqref{eq:dS-flow}, with $F_{\mathrm{dS}}$ replaced by $F_{\mathrm{AdS}}$, to the contour obtained by this continuation gives
\begin{align}
 \mathcal C_{\mathrm{AdS}}
 &\sim
 \mathcal J_0+\mathcal J_{-1}+\mathcal J_{-2}+\cdots,
 \label{eq:CAdS}
\end{align}
as shown in the right panel of Fig.~\ref{fig:bulk-thimbles}.
Here $\mathcal J_m$ denotes the Lefschetz thimble associated with the AdS saddle $N_m$.\footnote{As we will see, this contour is also consistent with holography.}
The coefficients in \eqref{eq:CAdS} are fixed by the continued dS result, not by an independent choice of AdS saddles.
The change from a single dS thimble to this infinite sum is the Stokes phenomenon.
At leading semiclassical order, we obtain\footnote{
Strictly speaking, the sum below does not converge for strictly real $\ell_{\mathrm{AdS}} /G_N$. 
The real-parameter limit is understood via analytic continuation, with the Gamma-function argument approaching the negative real axis from below.
}
\begin{align}
 Z_{\mathrm{AdS}}
 &\sim
 e^{\frac{\ell_{\mathrm{AdS}}}{4G_N}}
 \sum_{n=0}^{\infty}
 e^{\frac{2\pi i n\ell_{\mathrm{AdS}}}{4G_N}}.
 \label{eq:bulk-AdS}
\end{align}
These terms correspond to the smooth Euclidean AdS saddle and the family of Euclidean AdS geometries with additional bang/crunch regions found in \cite{Chen:2024qmn,Chen:2024vpa}.
Thus, analytic continuation from Lorentzian dS provides a contour prescription for the conformal factor problem in Euclidean AdS$_3$ minisuperspace.

%%%%%%%%%%%%%%%%%%%%%%%%%%%%%
%%%%%%%%%%%%%%%%%%%%%%%%%%%%%
%%%%%%%%%%%%%%%%%%%%%%%%%%%%%
\section{Liouville zero-mode}
%%%%%%%%%%%%%%%%%%%%%%%%%%%%%
%%%%%%%%%%%%%%%%%%%%%%%%%%%%%
%%%%%%%%%%%%%%%%%%%%%%%%%%%%%

We define the Liouville zero-mode integral on a single convergent Lefschetz thimble and analytically continue it to the AdS and dS branches.
This construction does not use the bulk lapse integral and therefore provides an independent test of the bulk saddle selection.

Let $g_{S^2}$ be the unit round metric,
\begin{align}
 \operatorname{Area}(g_{S^2})
 &=4\pi,
 &
 \int_{S^2}\!\sqrt{g_{S^2}}\,R_{g_{S^2}}
 &=8\pi.
\end{align}
We choose the spherical cutoff representative
\begin{align}
 g_\epsilon
 &=\frac{4}{\epsilon^2}g_{S^2}.
 \label{eq:cutoff}
\end{align}
We use $\epsilon_{\mathrm{AdS}}>0$ and $\epsilon_{\mathrm{dS}}>0$ to denote the dimensionless boundary cutoff parameters in AdS and dS, respectively.
To keep the associated UV cutoff length $\ell\epsilon$ fixed and positive, we use the AdS-to-dS continuation
\begin{align}
 \ell_{\mathrm{AdS}}
 &\longrightarrow i\ell_{\mathrm{dS}},
&
 \epsilon_{\mathrm{AdS}}
 &\longrightarrow-i\epsilon_{\mathrm{dS}}.
 \label{eq:eps}
\end{align}
Near the boundary, $\ell\epsilon\simeq\frac{\ell}{a_1}$, consistent with the bulk prescription.
The corresponding bulk continuation is $a_{1,\mathrm{AdS}}\to i a_{1,\mathrm{dS}}$.

On the cutoff geometry $g_\epsilon$, the Liouville action is
\begin{align}
 S_L[\phi;g_\epsilon]
 &=\frac1{4\pi}\int_{S^2}\!d^2x\sqrt{g_\epsilon}\,
 \left[(\nabla\phi)^2+Q R_{g_\epsilon}\phi
 +4\pi\mu e^{2b\phi}\right],
 \notag\\[-1mm]
 Q&=b+b^{-1}.
 \label{eq:SL}
\end{align}
The central charge determines $b$ through
\begin{align}
 c
 &=1+6Q^2
 =13+6\left(b^2+b^{-2}\right),
 \label{eq:central}
\end{align}
where we choose the branch satisfying $b^{-2}\sim c/6$ at large positive $c$.

%%%
\begin{figure*}[t]
 \centering
 \makebox[\textwidth][c]{%
  \includegraphics[width=0.3\textwidth]{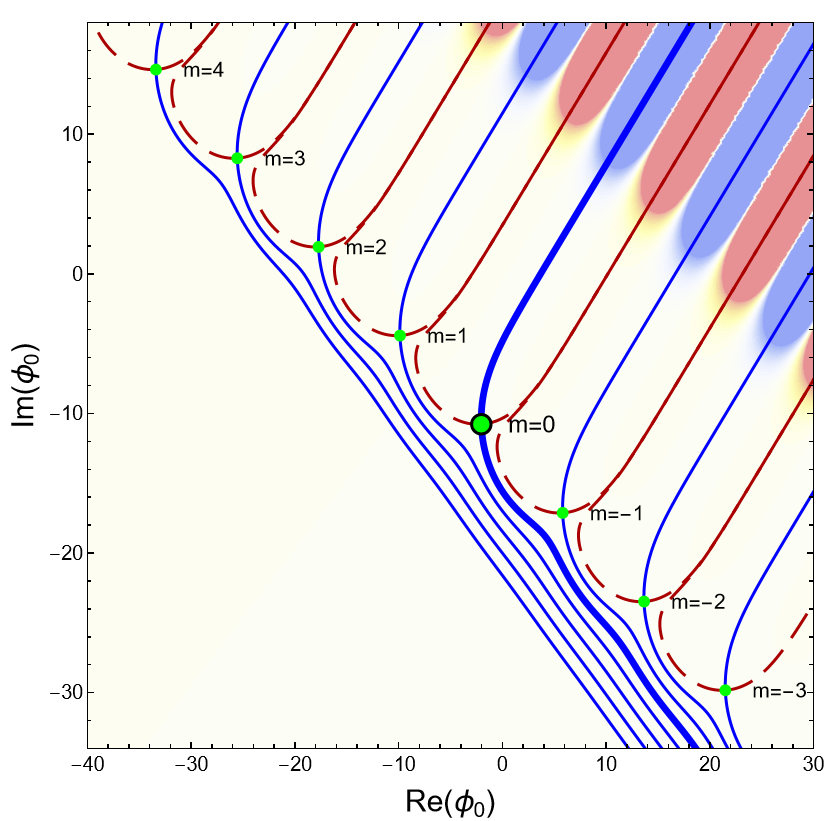}%
  \hspace{0.1\textwidth}%
  \includegraphics[width=0.3\textwidth]{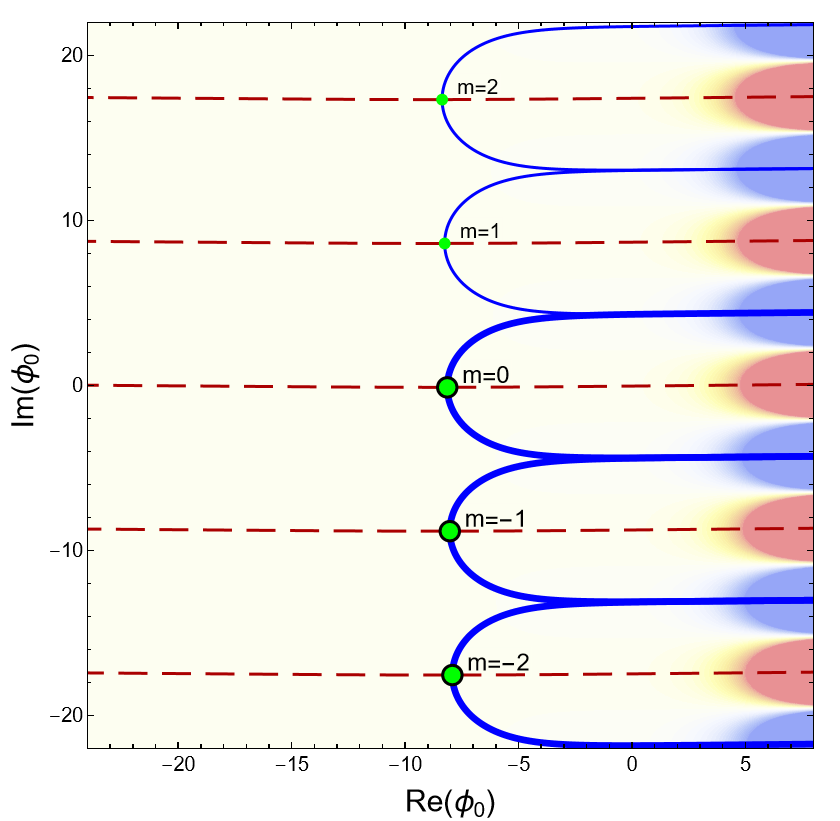}%
 }
 \caption{
Lefschetz-thimble decompositions in the complex $\phi_0$ plane for the Liouville zero-mode integral on the dS (left) and AdS (right) branches.
The solid blue curves, red dashed curves, and green markers denote the downward thimbles $\mathcal J_m$, upward cycles $\mathcal K_m$, and saddle points $\phi_m$, respectively.
Contributing thimbles and saddles are emphasized by thicker blue curves and larger green circles.
For both panels, we set $\epsilon_{\mathrm{AdS}}=\epsilon_{\mathrm{dS}}=0.1$ and $|c|=60$.
Left: for $c=60i$, the contour in \eqref{eq:y-natural} is homologous to $\mathcal J_0$, so only $m=0$ contributes.
Right: we take $b\simeq0.360245-0.004677i$, corresponding to $b^{-2}\simeq7.701645+0.2i$ and $c\simeq59.988400+1.179783i$ through \eqref{eq:central}.
Thus, $-1-b^{-2}$ lies below the negative real axis, and the continued contour is homologous to $\mathcal J_0+\mathcal J_{-1}+\mathcal J_{-2}+\cdots$.
The two panels exhibit the same transition from one contributing thimble in dS to infinitely many in AdS as in the bulk calculation in Fig.~\ref{fig:bulk-thimbles}.
}
 \label{fig:liouville-thimbles}
\end{figure*}
%%%

For the constant mode, the curvature term is fixed by the Gauss--Bonnet theorem, whereas the cosmological term contains the cutoff area:
\begin{align}
 \frac{Q}{4\pi}
 \int_{S^2}\!\sqrt{g_\epsilon}\,R_{g_\epsilon}\phi_0
 &=2Q\phi_0,
 \notag\\[-1mm]
 \frac1{4\pi}
 \int_{S^2}\!\sqrt{g_\epsilon}\,
 4\pi\mu e^{2b\phi_0}
 &=\frac{16\pi\mu}{\epsilon^2}e^{2b\phi_0}.
\end{align}
The resulting cutoff zero-mode path integral is
\begin{align}
 Z_{0,\epsilon}
 &=\int_{\mathcal C_\phi}\!d\phi_0\,
 e^{
 -2Q\phi_0
 -\frac{16\pi\mu}{\epsilon^2}e^{2b\phi_0}}.
 \label{eq:Z0-cutoff-integral}
\end{align}

To evaluate this integral, introduce $Y=\frac{16\pi\mu}{\epsilon^2}e^{2b\phi_0}$ on a fixed logarithmic sheet.
For $\operatorname{Re}(-\frac{Q}{b})>0$, the contour $Y\in\mathbb R_+$ is convergent and homologous to a single Lefschetz thimble.
Changing variables from $\phi_0$ to $Y$ then gives
\begin{align}
 Z_{0,\epsilon}
 &=\frac1{2b}
 \left(\frac{16\pi\mu}{\epsilon^2}\right)^{Q/b}
 \Gamma(-Q/b)
 \notag\\[-1mm]
 &=\frac1{2b}
 \left(\frac{16\pi\mu}{\epsilon^2}\right)^{1+b^{-2}}
 \Gamma(-1-b^{-2}).
 \label{eq:Z0}
\end{align}
This Gamma-function factor arises from the zero-mode integration in the full Liouville sphere partition function derived from the DOZZ formula \cite{Dorn:1994xn,Zamolodchikov:1995aa,Harlow:2011ny}.\footnote{For a discussion of the full Liouville sphere partition function, including the role of the conformal Killing group, see \cite{Giribet:2022cvw}.}
We use the analytic continuation of Eq.~\eqref{eq:Z0} to define $Z_{0,\epsilon}$ and its integration contour.
Equivalently, this contour can be obtained from the massless limit $\eta_i\to0$ at fixed $b$ of the Liouville zero-mode path integral with heavy insertions $e^{2\eta_i\phi_0/b}$.
We define this limit by analytically continuing the insertion charges from a convergent domain while deforming the integration contour, following the analytic-continuation prescription of \cite{Harlow:2011ny}.\footnote{
For a related analysis of the timelike Liouville zero mode on a prescribed Hankel-type contour, see \cite{Despontin:2026lvh}.}

The semiclassical normalization that reproduces the round sphere at leading order is
$\pi\mu b^2=-\frac14$.
Substituting this relation into \eqref{eq:Z0} gives the finite-cutoff semiclassical result
\begin{align}
 Z_{0,\epsilon}
 &=\frac1{2b}
 \left(-\frac{4}{b^2\epsilon^2}\right)^{1+b^{-2}}
 \Gamma(-1-b^{-2}).
 \label{eq:Z0-cutoff-Gamma}
\end{align}
All complex powers use this continued logarithmic branch.

We first consider the AdS branch.
For AdS, $\epsilon_{\mathrm{AdS}}^{\,2}>0$ and $b^{-2}>0$, so the base in \eqref{eq:Z0-cutoff-Gamma} lies on the negative real axis.
The argument of the Gamma function is
\begin{align}
 -1-b^{-2}
 &=-\frac{c}{6}+\frac76+\order(c^{-1}),
 \label{eq:zLE}
\end{align}
and therefore also lies on the negative real axis at large positive $c$.
Stirling's formula then gives the leading semiclassical expansion
\begin{align}
 Z_{0,\mathrm{AdS}}
 &\sim
 e^{\frac{c}{6}}
 % \left[1+\order(b^2)\right]
 \sum_{m=0}^{\infty}e^{\frac{2\pi i m c}{6}}.
 \label{eq:Liouville-AdS}
\end{align}
With the holographic dictionary $c=\frac{3\ell_{\mathrm{AdS}}}{2G_N}$, this reproduces the bulk result \eqref{eq:bulk-AdS} at leading semiclassical order.
The corresponding infinite-thimble decomposition is shown in the right panel of Fig.~\ref{fig:liouville-thimbles}.

We next continue to the dS branch by setting $c=ic_{\mathrm{dS}}$ and following the same branch of \eqref{eq:central}.
Then
\begin{align}
 -1-b^{-2}
 &=\frac76-\frac{i c_{\mathrm{dS}}}{6}
 +\order(c_{\mathrm{dS}}^{-1}).
 \label{eq:dS-b}
\end{align}
For sufficiently large $c_{\mathrm{dS}}$, 
the argument of the Gamma function has positive real part, so the integral converges on $Y\in\mathbb R_+$.
Under \eqref{eq:eps}, $\epsilon=-i\epsilon_{\mathrm{dS}}$, and this contour becomes
\begin{align}
 \frac{4}{b^2\epsilon_{\mathrm{dS}}^{\,2}}e^{2b\phi_0}
 &\in\mathbb R_+.
 \label{eq:y-natural}
\end{align}
This contour is homologous to a single Lefschetz thimble, as shown in the left panel of Fig.~\ref{fig:liouville-thimbles}.
The zero-mode partition function is therefore
\begin{align}
 Z_{0,\mathrm{dS}}
 &=
 \frac1{2b}
 \left(
 \frac{4}{b^2\epsilon_{\mathrm{dS}}^{\,2}}
 \right)^{1+b^{-2}}
 \Gamma(-1-b^{-2}).
 \label{eq:Z0-dS-cut}
\end{align}
Applying Stirling's formula gives
\begin{align}
 Z_{0,\mathrm{dS}}
 &\sim
 e^{-\frac{\pi c_{\mathrm{dS}}}{6}
 +\frac{i c_{\mathrm{dS}}}{6}}.
\end{align}
Thus, the Liouville zero-mode independently selects the same tunneling exponent.\footnote{At leading semiclassical order, this agrees with the weight of the $n=-1$ saddle in \cite{Hikida:2022ltr} (which is labeled as $m=0$ in the left panel of Fig.~\ref{fig:liouville-thimbles}).}

%%%%%%%%%%%%%%%%%%%%%%%%%%%%%
%%%%%%%%%%%%%%%%%%%%%%%%%%%%%
%%%%%%%%%%%%%%%%%%%%%%%%%%%%%
\section{Discussion}
%%%%%%%%%%%%%%%%%%%%%%%%%%%%%
%%%%%%%%%%%%%%%%%%%%%%%%%%%%%
%%%%%%%%%%%%%%%%%%%%%%%%%%%%%

Both the bulk minisuperspace path integral and the Liouville zero-mode integral reduce to Gamma functions and exhibit the same Lefschetz thimble structure.
In both calculations, the dS contour is homologous to a single Lefschetz thimble, yielding
\begin{align}
 |\Psi_{\mathrm{dS}}|
 &\sim
 |Z_{0,\mathrm{dS}}|
 \sim
 \exp\!\left[
 -\frac{\pi\ell_{\mathrm{dS}}}{4G_N}
 \right].
\end{align}
By contrast, in each calculation the analytically continued AdS contour is homologous to an infinite sum of Lefschetz thimbles.
Figures~\ref{fig:bulk-thimbles} and \ref{fig:liouville-thimbles} display this common transition from one contributing thimble in dS to infinitely many in AdS.
A related phenomenon appears in semiclassical Liouville theory on $\mathbb{RP}^2$, where a single saddle in the area variable is represented by an infinite tower of complex saddles in the logarithmic zero-mode variable \cite{Nakayama:2026wmp}.

The saddle weights agree at leading semiclassical order, but 
the arguments in the Gamma function differ at subleading order.
Assuming the holographic dictionary $c=\frac{3\ell_{\mathrm{AdS}}}{2G_N}$ at the semiclassical leading order, 
the arguments in the Gamma function for the AdS case are
\begin{align}
 \text{bulk:}\qquad
 &\frac14-\frac{c}{6},
 \notag\\
 \text{Liouville:}\qquad
 &\frac76-\frac{c}{6}+\order(c^{-1}),
\label{eq:dictionary-shift}
\end{align}
so their $\mathcal{O}(c^0 )$
%order-$c^0$ 
shifts differ by $\frac{11}{12}$.
Therefore, if this difference is absorbed entirely into the bulk–boundary parameter map, then imposing precise matching requires the following modification of the holographic dictionary: $\frac{3\ell_{\mathrm{AdS}}}{2G_N}=c-\frac{11}{2}+\order(c^{-1})$.
%Therefore imposing precise matching implies that the holographic dictionary should be modified as $\frac{3\ell_{\mathrm{AdS}}}{2G_N}=c-\frac{11}{2}+\order(c^{-1})$.
Similar modifications of holographic dictionaries beyond the semiclassical leading order have been mentioned in other holographic setups.
In ABJM theory, subleading corrections in the bulk shift the M2-brane charge and the AdS radius \cite{Bergman:2009zh,Aharony:2009fc}, while the $S^3$ partition function is naturally expressed in terms of a ``renormalized'' 't Hooft coupling \cite{Drukker:2010nc,Fuji:2011km,Marino:2011eh,Hanada:2012si}.
In $\mathcal{N}=4$ SYM, agreement of BPS state counting from the superconformal index and the supersymmetric AdS$_5$ black-hole entropy \cite{Cabo-Bizet:2018ehj,Choi:2018hmj,Benini:2018ywd} was found to persist formally at finite $N$ if we appropriately modify the relation between the Newton constant and $N$ \cite{Honda:2019cio}.
These examples suggest that holographic dictionaries at the quantum level may be modified from those at semiclassical leading order.
In our case, the $\frac{11}{12}$ difference in \eqref{eq:dictionary-shift} may be explained by such $\mathcal{O}(1)$ corrections to the bulk--boundary parameter map and/or to the reduced path-integral measures.

Closely related saddle families were obtained in \cite{Chen:2024qmn,Chen:2024vpa}, but the logic of the construction is different.
In those works, the semiclassical saddle content is first inferred from full Liouville correlators, and minisuperspace contours are then used to realize the resulting boundary-led answer.
In the dS analysis, in particular, the contour $N\in i\mathbb R_+$ associated with positive Lorentzian lapse is not adopted because it does not reproduce the two semiclassical CFT contributions corresponding to the tunneling and Hartle--Hawking weights.

The comparison also differs at the level of truncation.
Our two path integrals retain only the $SO(3)$-invariant variables: the bulk minisuperspace degree of freedom and the boundary Liouville zero-mode.
By contrast, \cite{Chen:2024qmn,Chen:2024vpa} use correlators of the full Liouville theory as boundary input while realizing the corresponding geometries in bulk minisuperspace.
The present agreement therefore provides a more controlled comparison between matched symmetry truncations, although it is not a more complete statement about the full theory.

This distinction also clarifies the different dS saddle content.
The single dS thimble found here is consistent with the Lorentzian minisuperspace analysis of \cite{Honda:2024aro, Honda:2026wsl} in several dimensions, in which the defining contour contains the tunneling saddles.\footnote{
In contrast to the pure Einstein gravity, a similar analysis in Jackiw-Teitelboim gravity shows that a Hartle--Hawking type saddle contributes to the path integral \cite{Honda:2024hdr}.
}
The difference should therefore be understood as a difference between truncations and nonperturbative contour prescriptions, rather than as a contradiction.
It remains possible that the additional Hartle--Hawking contribution is carried by nonzero Liouville modes or by integration cycles of the full theory.

Our construction also addresses the conformal factor problem in Euclidean AdS$_3$ minisuperspace by defining the path integral on a complex contour.
Already in this minisuperspace model, the naive integral over real Euclidean metrics does not converge.
A Euclidean saddle therefore does not by itself define the quantum theory.
It contributes only if its thimble appears with nonzero coefficient in the decomposition of the chosen integration contour.

Within the matched bulk and boundary truncations studied here, the agreement between the two calculations indicates that the choice of contour is part of the holographic prescription.
One should first choose a contour in a parameter region where the integral converges and then continue that contour as the parameters are varied.
The thimble decomposition may jump across Stokes lines, but once the initial contour is fixed, its continuation is not arbitrary.

\paragraph{Acknowledgments}
We are very grateful for valuable discussions with Yu Nakayama, Tadashi Takayanagi, and Yusuke Taki.
MH would like to thank Hiroki Matsui, Kota Numajiri, Kazumasa Okabayashi and Takahiro Terada for discussions on related works.
MH~is supported by JST CREST Grant Number JPMJCR24I3, JSPS Grant-in-Aid for Transformative Research Areas (A) ``Extreme Universe'' JP21H05190 [D01], JSPS KAKENHI Grant Number JP22H01222 and the Royal Society grants ICA/R2/242058 and IEC/R3/243103.
KS is supported by Grant-in-Aid for JSPS Fellows No.~25KJ1498. 

%%%%%%%%%%%%%%%%%%%%%%%%%%%%%
%%%%%%%%%%%%%%%%%%%%%%%%%%%%%
%%%%%%%%%%%%%%%%%%%%%%%%%%%%%
\bibliography{references}
%%%%%%%%%%%%%%%%%%%%%%%%%%%%%
%%%%%%%%%%%%%%%%%%%%%%%%%%%%%
%%%%%%%%%%%%%%%%%%%%%%%%%%%%%

\clearpage
\onecolumngrid
\begin{center}
 {\large\bfseries Supplemental Material for ``Stokes Phenomena between AdS/CFT and dS/CFT''}
\end{center}

\setcounter{section}{0}
\renewcommand{\thesection}{S\arabic{section}}
\renewcommand{\theHsection}{supp.\arabic{section}}
\setcounter{equation}{0}
\renewcommand{\theequation}{S\arabic{equation}}
\renewcommand{\theHequation}{supp.\arabic{equation}}
\setcounter{figure}{0}
\renewcommand{\thefigure}{S\arabic{figure}}
\renewcommand{\theHfigure}{supp.\arabic{figure}}

%%%%%%%%%%%%%%%%%%%%%%%%%%%%%
%%%%%%%%%%%%%%%%%%%%%%%%%%%%%
%%%%%%%%%%%%%%%%%%%%%%%%%%%%%
\section{Chern--Simons reduction and the Liouville sector}
\label{sec:CS-Liouville}
%%%%%%%%%%%%%%%%%%%%%%%%%%%%%
%%%%%%%%%%%%%%%%%%%%%%%%%%%%%
%%%%%%%%%%%%%%%%%%%%%%%%%%%%%

We summarize the classical reduction that motivates the Liouville
description used in the main text.
The reduction fixes the boundary field content and the leading
semiclassical dictionary, but it does not establish an exact equality
between the full gravitational and Liouville path integrals.
In particular, it does not determine their functional measures or
integration cycles.

%%%%%%%%%%%%%%%%%%%%%%%%%%%%%
%%%%%%%%%%%%%%%%%%%%%%%%%%%%%
\subsection{Chern--Simons formulation}
%%%%%%%%%%%%%%%%%%%%%%%%%%%%%
%%%%%%%%%%%%%%%%%%%%%%%%%%%%%

Let $e$ and $\omega$ denote the dreibein and the dualized spin
connection, viewed as $\mathfrak{sl}(2)$-valued one-forms.
For Lorentzian AdS$_3$ and dS$_3$, respectively, one introduces
\begin{align}
 A
 &=\omega+\frac{1}{\ell_{\mathrm{AdS}}}e,
&
 \bar A
 &=\omega-\frac{1}{\ell_{\mathrm{AdS}}}e,
&&
 (\text{Lorentzian AdS}_3),
\notag\\
 A
 &=\omega+\frac{i}{\ell_{\mathrm{dS}}}e,
&
 \bar A
 &=\omega-\frac{i}{\ell_{\mathrm{dS}}}e,
&&
 (\text{Lorentzian dS}_3).
\label{eq:CS-connections}
\end{align}
The first line defines two $SL(2,\mathbb R)$ connections, whereas the
second defines an $SL(2,\mathbb C)$ connection subject to a
gravitational reality condition.
Euclidean AdS$_3$ is described in the same complexified
$SL(2,\mathbb C)$ space with a different reality condition.
In each case, the flatness equations for $A$ and $\bar A$ are
equivalent to the torsion constraint and the Einstein equations with
the corresponding cosmological constant
\cite{Achucarro:1986uwr,Witten:1988hc,Cacciatori:2001un}.

In the standard normalization, define
\begin{align}
 I_{\rm CS}[A]
 &=
 \frac{1}{4\pi}
 \int_M
 \operatorname{Tr}\left(
 A\wedge dA+\frac{2}{3}A\wedge A\wedge A
 \right).
\label{eq:CS-action-supp}
\end{align}
For Lorentzian AdS$_3$, the gravitational action can be written, up to
boundary terms, as
\begin{align}
 I_{\rm grav}
 &=
 k_{\rm grav}
 \left(
 I_{\rm CS}[A]-I_{\rm CS}[\bar A]
 \right)
 +I_{\partial M},
\notag\\
 k_{\rm grav}
 &=
 \frac{\ell_{\mathrm{AdS}}}{4G_N}.
\label{eq:CS-level-supp}
\end{align}
The Brown--Henneaux central charge is therefore
\begin{align}
 c_{\rm BH}
 &=
 6k_{\rm grav}
 =
 \frac{3\ell_{\mathrm{AdS}}}{2G_N}.
\label{eq:BH-central-supp}
\end{align}
The Euclidean-AdS and Lorentzian-dS actions are obtained by
complexifying this expression and imposing the corresponding reality
condition.
Their overall phases depend on the signature and orientation, but the
flatness constraints and the local reduction described below have the
same complexified form.

%%%%%%%%%%%%%%%%%%%%%%%%%%%%%
%%%%%%%%%%%%%%%%%%%%%%%%%%%%%
\subsection{Boundary WZW model and Drinfeld--Sokolov reduction}
%%%%%%%%%%%%%%%%%%%%%%%%%%%%%
%%%%%%%%%%%%%%%%%%%%%%%%%%%%%

The boundary term $I_{\partial M}$ and the asymptotic boundary
conditions select a polarization for the Chern--Simons theory.
The components of the connections normal to a constant-radius slice
then impose the flatness constraints.
Locally near the boundary, these constraints are solved by
\begin{align}
 A
 &=
 G_1^{-1}dG_1,
&
 \bar A
 &=
 G_2^{-1}dG_2.
\label{eq:flat-CS-supp}
\end{align}
Substitution into the improved Chern--Simons action gives two chiral
WZW actions.
Using the Polyakov--Wiegmann identity, they combine into a nonchiral
WZW model for
\begin{align}
 g
 &=
 G_1^{-1}G_2,
\end{align}
schematically,
\begin{align}
 I_{\rm CS}[A]
 -I_{\rm CS}[\bar A]
 +I_{\partial M}
 \longrightarrow
 I_{\rm WZW}[g].
\label{eq:CS-WZW-supp}
\end{align}
This is the Chern--Simons/WZW step of the reduction
\cite{Witten:1988hf,Elitzur:1989nr,Coussaert:1995zp,
Banados:1998ta,Cacciatori:2001un}.
The statement is local in the boundary directions and does not by
itself classify global holonomy sectors.

Let $L_0,L_{\pm1}$ denote a standard basis of
$\mathfrak{sl}(2)$, and define the left- and right-moving currents by
\begin{align}
 J_{\bar z}
 &=
 g^{-1}\partial_{\bar z}g,
&
 \widetilde J_z
 &=
 \partial_zg\,g^{-1}.
\end{align}
Up to normalization and chirality conventions, the remaining
gravitational boundary conditions take the form
\begin{align}
 J_{\bar z}^{-}
 &=
 \lambda,
&
 \widetilde J_z^{+}
 &=
 \widetilde\lambda,
\notag\\
 J_{\bar z}^{0}
 &=
 0,
&
 \widetilde J_z^{0}
 &=
 0,
\label{eq:DS-constraints-supp}
\end{align}
where $\lambda$ and $\widetilde\lambda$ are nonzero constants.
The first line consists of first-class current constraints, while the
second line can be regarded as a gauge choice.
Together they implement the $\mathfrak{sl}(2)$
Drinfeld--Sokolov Hamiltonian reduction
\cite{Drinfeld:1984qv,Forgacs:1989ac,Balog:1990mu}.

To display the reduced action, use the Gauss decomposition
\begin{align}
 g
 &=
 \begin{pmatrix}
  1&X\\
  0&1
 \end{pmatrix}
 \begin{pmatrix}
   e^{\Phi/2}&0\\
  0& e^{-\Phi/2}
 \end{pmatrix}
 \begin{pmatrix}
  1&0\\
  Y&1
 \end{pmatrix}.
\label{eq:Gauss-WZW-supp}
\end{align}
Before imposing a reality condition, $X$, $Y$, and $\Phi$ are complex.
Up to the overall level and a signature-dependent phase, the WZW
action takes the local form
\begin{align}
 I_{\rm WZW}
 &=
 \frac{\kappa}{2\pi}
 \int d^2z\,
 \left[
 \frac12
 \partial_z\Phi\,\partial_{\bar z}\Phi
 +2 e^{-\Phi}
 \partial_{\bar z}X\,\partial_zY
 \right],
\label{eq:WZW-Gauss-supp}
\end{align}
where $\kappa$ is proportional to the Chern--Simons level.
The constraints in \eqref{eq:DS-constraints-supp} fix the momenta
conjugate to $X$ and $Y$.
The reduced action must therefore be obtained by a partial Legendre
transform rather than by directly substituting the constraints into
\eqref{eq:WZW-Gauss-supp}.
Eliminating $X$ and $Y$ then gives
\begin{align}
 I_{\rm red}
 &=
 \frac{\kappa}{2\pi}
 \int d^2z\,
 \left[
 \frac12
 \partial_z\Phi\,\partial_{\bar z}\Phi
 +2\Lambda_L e^\Phi
 \right],
\label{eq:Liouville-reduced-supp}
\end{align}
where $\Lambda_L$ is determined by
$\lambda\widetilde\lambda$ and may be changed by a constant shift of
$\Phi$.
After a field rescaling and a constant shift, this is the classical
Liouville action
\cite{Coussaert:1995zp,Forgacs:1989ac,Balog:1990mu,
Cacciatori:2001un}.

For Lorentzian dS$_3$, the gravitational reality condition used in
\cite{Cacciatori:2001un} leads to a real Liouville field with a
Euclidean Liouville action.
This classical statement is distinct from the quantum complex
Liouville proposal used in the main text
\cite{Hikida:2021ese,Hikida:2022ltr}.
Here the classical reduction is used only to motivate the Liouville
field and the leading semiclassical dictionary.
The complex central charge and the integration cycle are specified
separately by the complex Liouville proposal and by the analytic
continuation prescription of the main text.

%%%%%%%%%%%%%%%%%%%%%%%%%%%%%
%%%%%%%%%%%%%%%%%%%%%%%%%%%%%
\subsection{Semiclassical dictionary and matched truncations}
%%%%%%%%%%%%%%%%%%%%%%%%%%%%%
%%%%%%%%%%%%%%%%%%%%%%%%%%%%%

In the standard quantum normalization of Liouville theory
\cite{Dorn:1994xn,Zamolodchikov:1995aa}, the background charge and
central charge are
\begin{align}
 Q
 &=
 b+b^{-1},
&
 c
 &=
 1+6Q^2
 =
 13+6\left(b^2+b^{-2}\right).
\label{eq:Liouville-central-supp}
\end{align}
Combining the large-central-charge limit of
\eqref{eq:Liouville-central-supp} with
\eqref{eq:BH-central-supp} gives
\begin{align}
 b^{-2}
 &=
 \frac{c}{6}+\order(1)
 =
 k_{\rm grav}+\order(1)
 =
 \frac{\ell_{\mathrm{AdS}}}{4G_N}+\order(1).
\label{eq:b-k-dictionary-supp}
\end{align}
On the dS branch used in the main text,
\begin{align}
 c
 &=
 i c_{\mathrm{dS}},
&
 c_{\mathrm{dS}}
 &=
 \frac{3\ell_{\mathrm{dS}}}{2G_N},
\end{align}
and hence
\begin{align}
 b^{-2}
 &=
 i\frac{\ell_{\mathrm{dS}}}{4G_N}+\order(1).
\label{eq:b-dS-dictionary-supp}
\end{align}
Only the leading terms in
Eqs.~\eqref{eq:b-k-dictionary-supp} and
\eqref{eq:b-dS-dictionary-supp} follow from the classical reduction.
Determining the order-one terms requires the quantum reduction and its
measure.
The classical argument therefore does not predict the order-one
difference between the bulk and Liouville Gamma-function arguments
found in the main text.

The action of $SO(3)$ on the round $S^2$ is transitive.
Consequently, an $SO(3)$-invariant Liouville configuration satisfies
\begin{align}
 \phi(R\Omega)
 &=
 \phi(\Omega)
 \quad
 \text{for all }R\in SO(3)
 \qquad\Longrightarrow\qquad
 \phi(\Omega)=\phi_0.
\label{eq:SO3-zero-mode-supp}
\end{align}
Thus the Liouville constant mode is precisely the
$SO(3)$-invariant boundary sector.
After fixing radial diffeomorphisms, the metric ansatz
in \eqref{eq:metric} is the corresponding
$SO(3)$-invariant bulk sector.
The two calculations in the main text therefore use matched symmetry
truncations.

This identification does not equate the two reduced path integrals.
In particular, restricting the nonzero Liouville modes to zero is not
the same operation as integrating them out.
Neither the functional measures nor the integration cycles of the two
reduced theories are fixed by the classical Chern--Simons reduction.
Their agreement is therefore an independent semiclassical consistency
check rather than a consequence of an assumed equality.

%%%%%%%%%%%%%%%%%%%%%%%%%%%%%
%%%%%%%%%%%%%%%%%%%%%%%%%%%%%
\subsection{Cutoff dictionary}
%%%%%%%%%%%%%%%%%%%%%%%%%%%%%
%%%%%%%%%%%%%%%%%%%%%%%%%%%%%

We finally relate the bulk and boundary cutoff conventions.
At the AdS endpoint, the dimensionful boundary metric associated with
\eqref{eq:cutoff} is normalized as
\begin{align}
 h_{ij}^{\mathrm{AdS}}
 &=
 \ell_{\mathrm{AdS}}^2a_{1,\mathrm{AdS}}^{\,2}
 (g_{S^2})_{ij}
\notag\\
 &=
 \frac{\ell_{\mathrm{AdS}}^2}{4}
 (g_{\epsilon_{\mathrm{AdS}}})_{ij}
 =
 \frac{\ell_{\mathrm{AdS}}^2}{\epsilon_{\mathrm{AdS}}^{\,2}}
 (g_{S^2})_{ij}.
\label{eq:cutoff-metric-AdS-supp}
\end{align}
It follows that, at large cutoff,
\begin{align}
 a_{1,\mathrm{AdS}}^{\,2}
 &=
 \epsilon_{\mathrm{AdS}}^{-2}
 \left[
 1+\order(\epsilon_{\mathrm{AdS}}^{\,2})
 \right].
\label{eq:cutoff-AdS-supp}
\end{align}
In the spherical conformal representative of radius $\ell$, our UV cutoff length is $\frac{\ell}{a_1}$ \cite{Casini:2011kv}.
The cutoff dictionary gives $\frac{\ell}{a_1}=\ell\epsilon[1+\order(\epsilon^2)]$.

Under the continuation \eqref{eq:eps},
\begin{align}
 \ell_{\mathrm{AdS}}
 &\longrightarrow i \ell_{\mathrm{dS}},
&
 \epsilon_{\mathrm{AdS}}
 &\longrightarrow-i\epsilon_{\mathrm{dS}}.
\end{align}
The product $\ell\epsilon$ stays fixed and positive.
The induced boundary metric becomes complex along the path.
Since $a_1\sim\epsilon^{-1}$, the bulk cutoff transforms as $a_{1,\mathrm{AdS}}\to i a_{1,\mathrm{dS}}$.
Equivalently, $a_{1,\mathrm{dS}}\to-i a_{1,\mathrm{AdS}}$ and $a_{1,\mathrm{dS}}^{\,2}\to-a_{1,\mathrm{AdS}}^{\,2}$, as in \eqref{eq:analyticcont}.

Writing the final metric in terms of the positive dS cutoff parameter
$\epsilon_{\mathrm{dS}}$ gives
\begin{align}
 h_{ij}^{\mathrm{dS}}
 &=
 \ell_{\mathrm{dS}}^2a_{1,\mathrm{dS}}^{\,2}
 (g_{S^2})_{ij}
\notag\\
 &=
 \frac{\ell_{\mathrm{dS}}^2}{4}
 (g_{\epsilon_{\mathrm{dS}}})_{ij}
 =
 \frac{\ell_{\mathrm{dS}}^2}{\epsilon_{\mathrm{dS}}^{\,2}}
 (g_{S^2})_{ij}.
\label{eq:cutoff-metric-dS-supp}
\end{align}
Thus
\begin{align}
 a_{1,\mathrm{dS}}^{\,2}
 &=
 \epsilon_{\mathrm{dS}}^{-2}
 \left[
 1+\order(\epsilon_{\mathrm{dS}}^{\,2})
 \right].
\label{eq:cutoff-dS-supp}
\end{align}

Combining the cutoff relations with
Eqs.~\eqref{eq:b-k-dictionary-supp} and
\eqref{eq:b-dS-dictionary-supp} gives
\begin{align}
 -\frac{4}{b^2\epsilon_{\mathrm{AdS}}^{\,2}}
 &=
 -4\frac{\ell_{\mathrm{AdS}}}{4G_N}
 a_{1,\mathrm{AdS}}^{\,2}
 \left[
 1+\order\left(\frac{G_N}{\ell_{\mathrm{AdS}}}\right)
 +\order(\epsilon_{\mathrm{AdS}}^{\,2})
 \right],
\notag\\
 \frac{4}{b^2\epsilon_{\mathrm{dS}}^{\,2}}
 &=
 4i\frac{\ell_{\mathrm{dS}}}{4G_N}
 a_{1,\mathrm{dS}}^{\,2}
 \left[
 1+\order\left(\frac{G_N}{\ell_{\mathrm{dS}}}\right)
 +\order(\epsilon_{\mathrm{dS}}^{\,2})
 \right].
\label{eq:Gamma-base-match-supp}
\end{align}
These are the leading semiclassical relations between the
finite-cutoff Gamma-function bases in the bulk and boundary
calculations.
The relative sign follows from
$\epsilon_{\mathrm{AdS}}^{\,2}\to-\epsilon_{\mathrm{dS}}^{\,2}$.
The Chern--Simons reduction motivates these variables and fixes their
leading dictionary, but it does not impose the Lefschetz thimble
decompositions found independently in the main text.

%%%%%%%%%%%%%%%%%%%%%%%%%%%%%
%%%%%%%%%%%%%%%%%%%%%%%%%%%%%
%%%%%%%%%%%%%%%%%%%%%%%%%%%%%
\section{Exact evaluation of the finite-cutoff minisuperspace integral}
%%%%%%%%%%%%%%%%%%%%%%%%%%%%%
%%%%%%%%%%%%%%%%%%%%%%%%%%%%%
%%%%%%%%%%%%%%%%%%%%%%%%%%%%%
The dS minisuperspace integral \eqref{eq:Psi-lapse} admits a closed expression at finite cutoff as follows.
We first take the limit $\varepsilon\to0^+$ in \eqref{eq:Psi-lapse} at fixed $a_1>0$ and introduce
\begin{equation}
u=\frac{1}{e^{-2iN}-1},
\qquad
N=\frac{i}{2}\log\frac{1+u}{u}.
\end{equation}
Using the counterterm $I_{\rm ct}^{\mathrm{dS}}=-2i\frac{\ell_{\mathrm{dS}}}{4G_N}a_1^2$ in the main text, we obtain
\begin{align}
\Psi_{\mathrm{dS}}
&=
\frac{e^{i\pi/4}}{\sqrt{2}}
\int_0^\infty du\,
u^{-\frac34-\frac{i\ell_{\mathrm{dS}}}{4G_N}}
(1+u)^{-\frac34+\frac{i\ell_{\mathrm{dS}}}{4G_N}}
e^{-\frac{i\ell_{\mathrm{dS}}}{G_N}a_1^2u}
\nonumber\\
&=
\frac{e^{i\pi/4}}{\sqrt{2}}\,
\Gamma\left(
\frac14-\frac{i\ell_{\mathrm{dS}}}{4G_N}
\right)
U\left(
\frac14-\frac{i\ell_{\mathrm{dS}}}{4G_N},
\frac12,
\frac{i\ell_{\mathrm{dS}}}{G_N}a_1^2
\right).
\label{eq:ds-closed}
\end{align}
where $U(a,b,z)$ is Tricomi's confluent hypergeometric function  defined as
\begin{equation}
U(a,b,z)
=
\frac{1}{\Gamma(a)}
\int_0^\infty dt\,
e^{-zt} t^{a-1}(1+t)^{b-a-1},
\qquad
\left( \operatorname{Re}a>0,\ \operatorname{Re}z>0 \right) .
\label{eq:Trocomi}
\end{equation}
The integral in \eqref{eq:ds-closed} is absolutely convergent at both endpoints, so the representation \eqref{eq:Trocomi} extends to these parameters by continuity from $\operatorname{Re}z>0$.
Taking $a_1\to\infty$ at fixed $\ell_{\mathrm{dS}}/G_N$ and using the asymptotic formula
\begin{equation}
U(a,b,z)
=
z^{-a}\left[1+O(z^{-1})\right] ,
\end{equation}
we find
\begin{equation}
\Psi_{\mathrm{dS}}
=
\frac{e^{i\pi/4}}{\sqrt{2}}\,
\Gamma\left(
\frac14-\frac{i\ell_{\mathrm{dS}}}{4G_N}
\right)
\left(
\frac{i\ell_{\mathrm{dS}}}{G_N}a_1^2
\right)^{\frac{i\ell_{\mathrm{dS}}}{4G_N}-\frac14}
\left[1+O(a_1^{-2})\right],
\end{equation}
which reproduces \eqref{eq:Psi-Gamma-cutoff}.
For fixed $0<\varepsilon<\pi/2$, deforming the regulated contour $-\varepsilon+i\mathbb{R}_+$ to $i\mathbb{R}_+$ produces an endpoint contribution of order $e^{-\ell_{\mathrm{dS}}a_1^2\cot\varepsilon/(2G_N)}$
as $a_1\to\infty$.
Thus, \eqref{eq:Psi-Gamma-cutoff} also holds when $a_1\to\infty$ is taken before
$\varepsilon\to0^+$, as prescribed in the main text.

%%%%%%%%%%%%%%%%%%%%%%%%%%%%%
%%%%%%%%%%%%%%%%%%%%%%%%%%%%%
%%%%%%%%%%%%%%%%%%%%%%%%%%%%%
\section{Stokes phenomenon of the Gamma function}
%%%%%%%%%%%%%%%%%%%%%%%%%%%%%
%%%%%%%%%%%%%%%%%%%%%%%%%%%%%
%%%%%%%%%%%%%%%%%%%%%%%%%%%%%

The Gamma function is the elementary example of the Stokes phenomenon used in the main text,
\begin{align}
 \Gamma(z)
 &=\int_0^\infty\!dY\,Y^{z-1} e^{-Y}
 \notag\\[-1mm]
 &=\int_{-\infty}^{\infty}\!dW\,
  e^{zW- e^W},
 \qquad \operatorname{Re}z>0,
 \label{eq:Gamma}
\end{align}
where $W=\log Y$. The exponent
\begin{align}
 F_\Gamma(W)
 &=zW- e^W
\end{align}
has the saddles
\begin{align}
 W_m
 &=\Log z+2\pi i m,
 \qquad m\in\mathbb Z,
 \label{eq:Gamma-saddles}
\end{align}
with action differences $F_\Gamma(W_m)-F_\Gamma(W_0)=2\pi i m z$. The Stokes condition between two such saddles is therefore $\operatorname{Re}z=0$. For $\operatorname{Re}z>0$, the contour $Y\in\mathbb R_+$, equivalently $W\in\mathbb R$, is homologous to one thimble. Continuing $z$ across $\operatorname{Re}z=0$ changes this decomposition, and the continued contour in the left half-plane is homologous to an infinite sum of thimbles.

\begin{figure*}[t]
\begin{center}
 \includegraphics[width=0.30\textwidth]{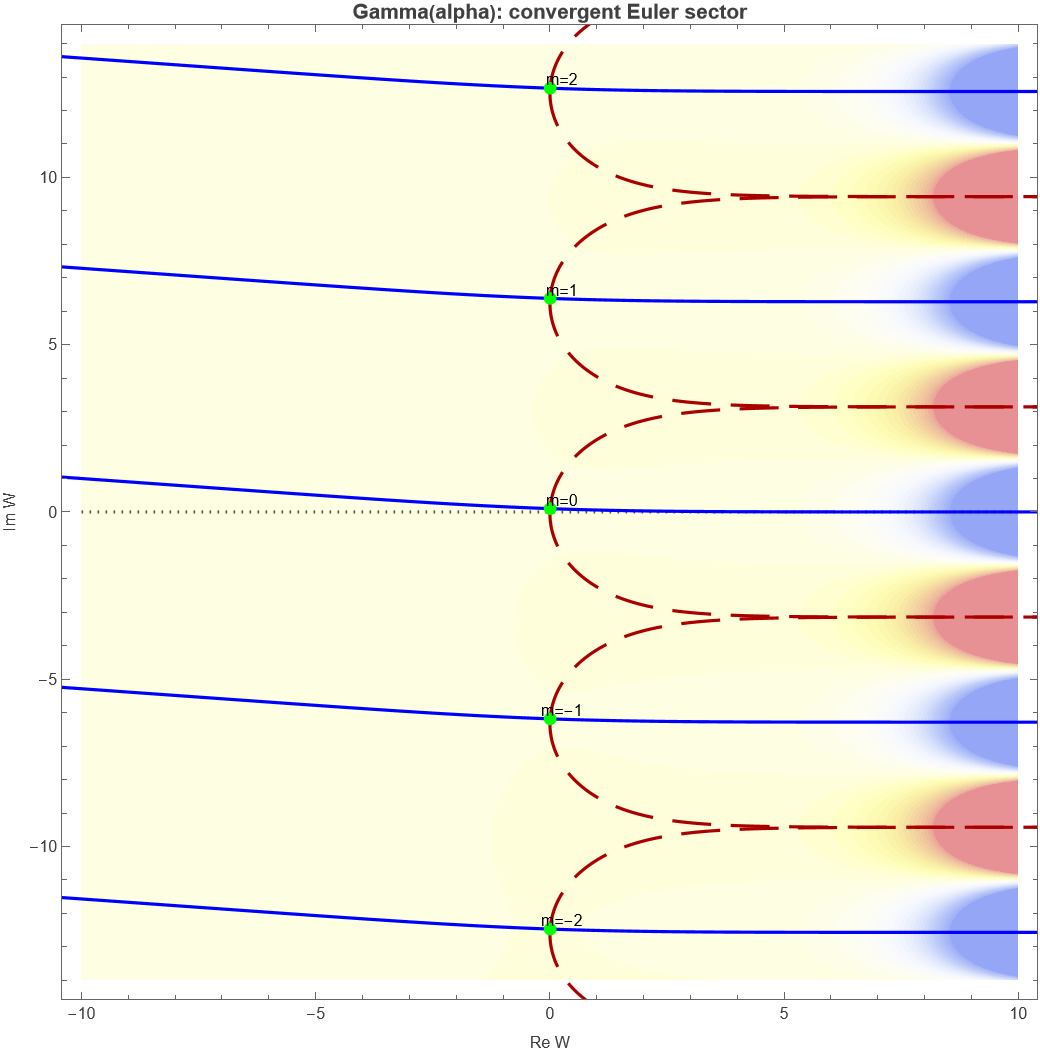}%
 \hspace{0.05\textwidth}%
 \includegraphics[width=0.30\textwidth]{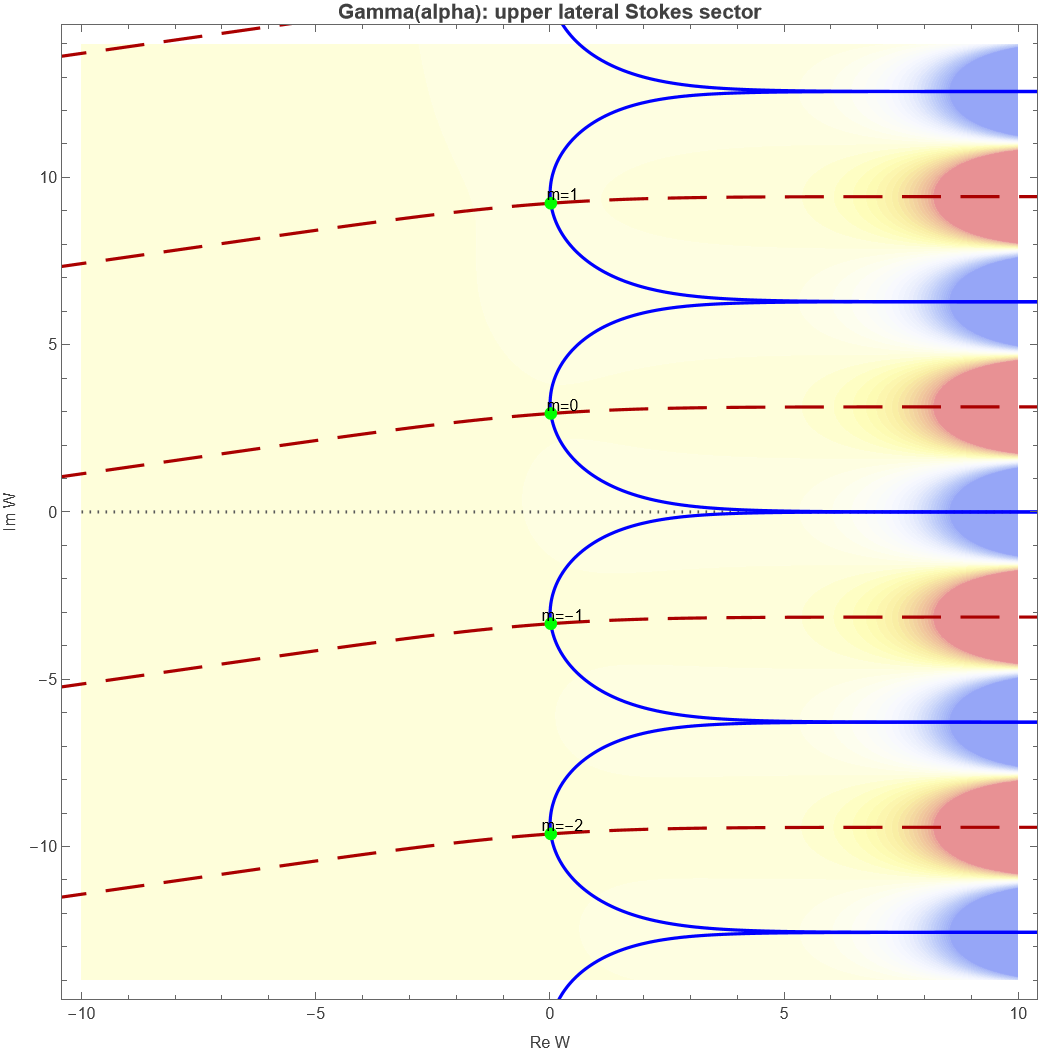}
 \captionof{figure}{Gradient-flow geometry for $F_\Gamma(W)=zW- e^W$. Blue curves are downward thimbles $\mathcal J _m$, red dashed curves are upward cycles $\mathcal K _m$, and green points are the saddles $W_m=\Log z+2\pi i m$. Left: for $\operatorname{Re}z>0$, the contour $Y\in\mathbb R_+$ is homologous to one thimble. Right: for $\operatorname{Re}z<0$ and $\operatorname{Im}z>0$, its analytic continuation is homologous to an infinite sum of thimbles. The geometry for $\operatorname{Im}z<0$ is obtained by reflection across the real $W$ axis.}
 \label{fig:gamma-stokes}
\end{center}
\end{figure*}

Suppressing the inverse-power Stirling series common to every saddle, the large-$|z|$ expansions are
\begin{align}
 \Gamma(z)
 &\sim\sqrt{2\pi}\,z^{z-\frac12} e^{-z},
 &&\operatorname{Re}z>0,
 \notag\\[-1mm]
 \Gamma(z)
 &\sim\sqrt{2\pi}\,z^{z-\frac12} e^{-z}
 \sum_{m=0}^{\infty} e^{2\pi i m z},
 &&\operatorname{Re}z<0,
 \quad \operatorname{Im}z>0,
 \notag\\[-1mm]
 \Gamma(z)
 &\sim\sqrt{2\pi}\,z^{z-\frac12} e^{-z}
 \sum_{m=0}^{\infty} e^{-2\pi i m z},
 &&\operatorname{Re}z<0,
 \quad \operatorname{Im}z<0.
 \label{eq:Gamma-Stokes}
\end{align}
In each line, $z^{z-1/2}$ is defined with the branch of $\Log z$ obtained by continuation along the specified path. The second and third lines describe continuations through the upper and lower half-planes, respectively. Their different sums of exponentially shifted saddle contributions are the Gamma-function Stokes phenomenon encoded in the exponentially improved Stirling expansion \cite{Nemes:2015gamma}. The dS-to-AdS continuation in the main text reaches the negative real axis from $\operatorname{Im}z<0$ and therefore uses the last line of  \eqref{eq:Gamma-Stokes}. The right panel of Fig.~\ref{fig:gamma-stokes} shows the reflected geometry with $\operatorname{Im}z>0$.

%%%
%%%

\end{document}